\documentclass[twocolumn,prl,tightenlines,superscriptaddress,showpacs,longbibliography]{revtex4-1}
\usepackage{amsmath}
\usepackage{amssymb,amsfonts,latexsym}
\usepackage{bm}
\usepackage[mathcal]{euscript}
\usepackage{graphicx}
\usepackage{epsfig}
\usepackage{color}
\usepackage{xfrac}
\usepackage{hyperref}  
\usepackage{url}
\usepackage{footmisc}

\begin{document}

\title{Quantitative Assessment of the Toner and Tu Theory of Polar Flocks}

\author{Beno\^it Mahault}
\affiliation{Service de Physique de l'Etat Condens\'e, CEA, CNRS, Universit\'e Paris-Saclay, CEA-Saclay, 91191 Gif-sur-Yvette, France}
\affiliation{Max Planck Institute for Dynamics and Self-Organization (MPIDS), 37077 G\"ottingen, Germany}

\author{Francesco Ginelli}
\affiliation{Department of Physics and Institute for Complex Systems and Mathematical Biology, Kings College, University of Aberdeen, Aberdeen AB24 3UE, United Kingdom}

\author{Hugues Chat\'{e}}
\affiliation{Service de Physique de l'Etat Condens\'e, CEA, CNRS, Universit\'e Paris-Saclay, CEA-Saclay, 91191 Gif-sur-Yvette, France}
\affiliation{Beijing Computational Science Research Center, Beijing 100094, China}
\affiliation{LPTMC, CNRS UMR 7600, Universit\'e Pierre et Marie Curie, 75252 Paris, France}

\date{\today}
%\pacs{05.65.+b, 45.70.Vn, 87.18.Gh}

\begin{abstract}
We present a quantitative assessment of the Toner and Tu theory describing the universal 
scaling of fluctuations in polar phases of dry active matter. 
Using large scale simulations of the Vicsek model in two and three dimensions, 
we find the overall phenomenology and generic algebraic scaling predicted by Toner and Tu, but 
our data on density correlations reveal some qualitative discrepancies.
The values of the associated scaling exponents we estimate differ significantly from those conjectured in 1995. 
In particular, we identify a large crossover scale beyond which flocks are only weakly anisotropic. 
We discuss the meaning and consequences of these results.
\end{abstract}

\maketitle

%%%%%%%%% Introduction %%%%%%%%%

Two seminal papers, both published in this journal in 1995, 
can be argued to mark the birth of active matter physics.
In \cite{vicsek1995novel}, Vicsek and collaborators introduced their simple model for collective motion, 
where XY spins fly at constant speed along their magnetic direction.
In \cite{TT95}, Toner and Tu (TT) wrote down fluctuating hydrodynamic equations for this flying XY model 
and performed a dynamic renormalization group calculation of its ordered phase, concluding, among other things, 
that such polar flocks possess true long-range orientational order even in two space dimensions (2D). 
In other words, flying spins defy the famous Mermin-Wagner theorem~\cite{Mermin1966}.
Since then, our knowledge of active matter has expanded tremendously (see, e.g., the various review papers 
\cite{toner2005hydrodynamics,ramaswamy2010review,Romanczuk2012ABP,vicsek2012collective,Marchetti2013RMP,cates2015MIPS,Elgeti2015RMP,prost2015activegels,bechinger2016ABP,Ramaswamy2017activematter,Doostmohammadi2018AN,chate-AR2019}). 
But the TT papers remain  influential
even though they deal with the limit case of dilute, aligning, dry active matter, 
which usually consists of self-propelled particles
subjected to local alignment in the absence of any surrounding fluid~\cite{chate-AR2019}.
In particular, the TT theory (and related works by Ramaswamy {\it et al.}) 
predicted what has become one of the most popular features in active matter studies, 
the presence, in orientationally-ordered phases, of ``giant number fluctuations" 
where the variance of the number of particles in sub-systems of 
increasing size scales faster than the mean
\cite{Ramaswamy2003GNF,chate2006Active_nema,Narayan2007GNF,chate2008collective,ginelli2010rods,zhang2010collective,ngo2014AN,Giavazzi2017GNF,nishiguchi2017LRO}.

Over the years, numerous numerical and experimental works have tried
to verify the TT results, but the evidence presented has been restricted
to a limited range of scales \cite{toner1998soundwaves} and/or isotropic measures
averaged over all spatial directions that cannot resolve individual
scaling exponents \cite{chate2008collective,chate2008modeling,Giavazzi2017GNF},
resulting in exponent values that could only be deemed compatible with the TT predictions.
This situation was satisfactory as long as the TT theory was believed, as claimed in the early papers
\cite{TT95,TT98}, to be `exact at all orders' in 2D, the dimension of choice of most works.
However, Toner himself realized in 2012 \cite{toner2012reanalysis} that this is not actually true, 
and that a number of important terms has been overlooked, invalidating most claims of exactness. 
The remarkable result of true long-range order in 2D remains valid, as well as the overall structure of the theory,
but scaling exponent values, and other important features, had to be `revisited'. 
From then on, belief in the TT results became reliant on the partial numerical evidence mentioned above. 
In spite of this situation, not much further work was devoted to gauge the accuracy of the TT predictions 
(see however \cite{Kyriakopoulos2016NJP,geyer2018sounds}), and a full-fledged, quantitative evaluation of the TT theory is still missing.

In this Letter, we present large-scale numerical simulations of the Vicsek model designed to study
the 2D and 3D anisotropic space-time correlations functions at the
heart of TT theory. 
Our results largely confirm its qualitative validity, 
but our estimates of exponent values clearly differ from the conjectured ones. 
In particular, we find that anisotropy is weak, possibly vanishing. 
Moreover, the behavior of density correlations shows qualitative discrepancies with the theory.
We discuss their origin, as well as the theoretical consequences of the hyperscaling relations that we find numerically satisfied.
 
%%%%%%%%% Summary of Toner and Tu's results %%%%%%%%%

We start with a synthetic account of the TT theory.
The hydrodynamic equations written by Toner and Tu govern
a conserved density $\rho$ and a velocity field ${\bf v}$:  
\begin{subequations}
\begin{align}
\label{eq:Toner_Tu_rho} & \partial_t \rho + \nabla \cdot (\rho {\bf v}) = 0 \,, \\
\label{eq:Toner_Tu_v} & \partial_t {\bf v}
\!+\! \lambda_1 ({\bf v}  \!\cdot\! \nabla){\bf v} \!+\! \lambda_2 (\nabla\! \cdot \!{\bf v} ){\bf v}  \!+\! \lambda_3 \nabla |{\bf v}|^2 
\!=\! \left[\alpha \!-\! \beta |{\bf v}|^2\right] \! {\bf v} \nonumber \\
&- \nabla P + D_0 \nabla^2 {\bf
  v} + D_1 \nabla (\nabla \!\cdot\! {\bf v} ) + D_2 ({\bf v} \!\cdot\! \nabla)^2 {\bf v}  +  {\bf f} 
\end{align}
\label{eq:Toner_Tu}
\end{subequations}
Here all coefficients can in principle depend on $\rho$ and $|{\bf v}|$, 
the pressure $P$ is expressed as a series in the density,
and ${\bf f}$ is an additive noise with zero mean and variance $\Sigma$ delta-correlated in space and time.
To obtain the quantities of interest hereafter, i.e. correlation functions of density and transverse velocity fluctuations,
Eqs.~(\ref{eq:Toner_Tu}) are linearized around the homogeneous ordered solution: $\rho = \rho_0 + \delta\rho$ and ${\bf v} = \left( v_0 + \delta v_\|\right) \hat{{\bf e}}_\| + \delta{\bf v}_\perp$, with $\rho_0$ the global density and  $v_0 = \sqrt{\alpha/\beta}$.
(Hereafter subscripts $_\|$ and $_\perp$ refer respectively to directions longitudinal and transverse to global order.)
After enslaving the fast field $\delta v_\|$, the Fourier-transformed
slow fluctuations read, 
in the small $q = |{\bf q}|$ limit \cite{toner2012reanalysis}:
\begin{subequations}
\begin{align}
\label{eq:Space-time_CF_rho} & \langle | \delta \rho(\omega,{\bf q}) |^2 \rangle = \frac{\rho_0^2 \Sigma}{{\cal S}(\omega,{\bf q})}q_\perp^2 \,, \\
\label{eq:Space-time_CF_v} &  \langle | \delta {\bf v}_\perp(\omega,{\bf q}) |^2 \rangle = \frac{\Sigma (\omega - v_2 q_\|)^2}{{\cal S}(\omega,{\bf q})} +
 \frac{\Sigma (d-2)}{{\cal S}_{\rm T}(\omega,{\bf q})} \,,
\end{align}
\label{eq:Space-time_CF}
\end{subequations}
where
${\cal S}(\omega,{\bf q}) = [(\omega - c_+(\theta_{\bf q})q)^2 + \varepsilon_+^2({\bf q})][(\omega - c_-(\theta_{\bf q})q)^2 + \varepsilon_-^2({\bf q})]$,
${\cal S}_{\rm T}(\omega,{\bf q}) = (\omega - c_{\rm T}(\theta_{\bf q})q)^2 + \varepsilon_{\rm T}^2({\bf q})$, with
$\theta_{\bf q}$ the angle between global order
and ${\bf q}$, $q_\perp = |{\bf q}_\perp|$.
The definitions of $v_2$, $\gamma$, $c_{\pm,{\rm T}}(\theta_{\bf q})$, $\varepsilon_{\pm,{\rm T}}({\bf q})$,
which are unimportant for the following discussion,
can be found in~\cite{toner2012reanalysis}.

Eq.~(\ref{eq:Space-time_CF_rho}) implies the existence of propagative sound modes, or density waves, 
whose dispersion relations follow 
$\omega_\pm({\bf q}) = c_\pm(\theta_{\bf q})q - \imath \varepsilon_\pm({\bf q})$.
This endows density fluctuations $\langle | \delta \rho(\omega,{\bf q}) |^2 \rangle$  with two sharp peaks in $\omega$ 
centered in $c_\pm(\theta_{\bf q})q$ and of respective widths $\varepsilon_\pm({\bf q})$.
The two terms of the rhs of Eq.~(\ref{eq:Space-time_CF_v}) correspond respectively to
transverse velocity fluctuations parallel and perpendicular to ${\bf q}_\perp$. 
The first term represents correlations of $v_{\rm L} =  \delta {\bf v}_\perp \cdot \hat{{\bf q}}_\perp$, which behave
like the density fluctuations.
The second term denotes the fluctuations of ${\bf v}_{\rm T} =  \delta {\bf v}_\perp - v_{\rm L}\hat{{\bf q}}_\perp$, which exist only for $d > 2$, and yields  a third peak centered in $c_{\rm T}(\theta_{\bf q})q$, of width $\varepsilon_{\rm T}({\bf q})$.

Since $\varepsilon_{\pm,{\rm T}}({\bf q})$ essentially scale as $q^2$ in the small wavenumber limit \cite{toner2012reanalysis},
the equal-time correlation functions are easily obtained by integrating Eqs.~(\ref{eq:Space-time_CF}) 
over $\omega$.
The resulting expressions, presented in \cite{toner2012reanalysis}, 
imply that $\langle | \delta {\bf v}_\perp({\bf q}) |^2 \rangle \approx q^{-2}$ when $q\to 0$,
whose primary consequence is the absence of long-range order in $d\le2$. 

However, nonlinearities in Eqs.~(\ref{eq:Toner_Tu}) are relevant perturbations 
for all $d \le d_c = 4$~\cite{toner2012reanalysis}.
Correlation functions in the nonlinear theory are then given by Eqs.~(\ref{eq:Space-time_CF}) using the 
renormalized noise variance and sound modes dampings
\begin{equation}
\Sigma^* = q_\perp^{z - \zeta} \, f_\Sigma(q_\|/q_\perp^\xi), \;\;
\varepsilon^*_{\pm,{\rm T}} = q_\perp^{z} \,f_{\pm,{\rm T}}(q_\|/q_\perp^\xi) , \label{eq:scaling_damplings}
\end{equation}
with $\zeta\equiv d-1+2\chi+ \xi$,
$f_{\Sigma,\pm,{\rm T}}(x) = {\cal O}(1)$ for $x\to 0$,
$f_{\Sigma}(x) \sim x^{(z-\zeta)/\xi}$ and $f_{\pm,{\rm T}}(x) \sim x^{z/\xi}$ when $x\to \infty$,
while the sound speeds $c_{\pm,{\rm T}}(\theta_{\bf q})$ remain those given by the linear theory.

%%%%%%%%%%%%%%%%%%%%%%%%%%%
\begin{table}[t!]
	\centering
\caption{Exponent values conjectured by Toner and Tu in~\cite{TT95} and those
resulting from our numerical evaluation of the density and velocity correlation functions.}
\begin{ruledtabular}
\begin{tabular}{lccccc}
		& \multicolumn{2}{c}{$d = 2$}  & \multicolumn{2}{c}{$d=3$} & $d \ge 4$\\
		& TT95 & numerics & TT95 & numerics & mean-field\\	
		$\chi$ & $-0.20$ & $-0.31(2)$ & $-0.60$ & $\simeq -0.62$ & $1-d/2$\\
		$\xi$ & $0.60$ & $0.95(2)$ & $0.80$ & $\simeq 1$ & $1$\\
		$\zeta=d\!-\!1 \!+\! 2\chi \!+\! \xi$ & $1.20$ & $1.33(2)$ & 1.60 & $1.77(3)$ & $2$ \\
		$z$ & $1.20$ & $1.33(2)$ & $1.60$ & $\simeq 1.77$ & $2$\\
		GNF & $1.60$ & $1.67(2)$ & $1.53$ & $1.59(3)$ & $1 + 2/d$\\
 	\end{tabular}
	\end{ruledtabular}
	\label{tab:Exponents}
\end{table}
%%%%%%%%%%%%%%%%%%%%%%%%%%%

Exponents $\chi$, $\xi$ and $z$ and scaling functions $f_{\Sigma,\pm,{\rm T}}$ are universal. 
The roughness exponent $\chi$ rules how the variance of velocity and density fluctuations varies with lengthscales.
Fluctuations vanish asymptotically when $\chi < 0$, insuring long-range polar order. Toner and Tu's calculations proved that
this is true for $d=2$ and $3$, while in linear theory, where $\chi=1-d/2$, fluctuations diverge and order is destroyed in 2D.
The anisotropy exponent $\xi$ measures the difference in scaling along and transversally to global order.
TT theory predicts that fluctuations scale anisotropically for
$d<d_c=4$ ($\xi<1$) while
 in mean-field $\xi=1$.
Finally, the dynamical exponent $z$ measures how the lifetime of sound modes scales with system size.
At the linear level $z=2$, which corresponds to a diffusive damping, while $z < 2$ is expected for $d < 4$ according to TT theory.
In their first publications \cite{TT95,TT98},
 Toner and Tu claimed an exact computation of these exponents in $d=2$, and found
 $\chi=(3-2d)/5$, $\xi=z/2=(d+1)/5$ (see TT95 numbers in Table~\ref{tab:Exponents}).
In his later ``reanalysis'' of the theory~\cite{toner2012reanalysis}, 
Toner realized that additional relevant nonlinearities were missed,
so that the above exponent values could only be exact, even in $d=2$, under the conjecture
of the asymptotic irrelevance of these terms.

We now turn to our numerical assessment of TT theory. We use the standard discrete-time Vicsek model for
efficiency. 
Particles $i=1,\ldots, N$ with position ${\bf r}_i$ and orientation ${\bf \hat{e}}_i$ move at constant speed $v_0$ and align
their velocities with current neighbors $j$:
\begin{equation}
{\bf \hat{e}}_i^{t+1} =  \vartheta \left[ 
\langle {\bf \hat{e}}_j^{t} \rangle_{j\sim i} + \eta \text{\boldmath$\xi$}^t_i \right]\,, \;\;
{\bf r}_i^{t+1} = {\bf r}_i^t + v_0 {\bf \hat{e}}_i^{t+1} \,,
\label{eq:micro_model}
\end{equation}
where $\vartheta[{\bf u}] = {\bf u}/|{\bf u}|$,
$\langle . \rangle_{j\sim i}$ is the average over all particles $j$ within unit distance of $i$ (including $i$), 
and $\text{\boldmath$\xi$}_i^t$ are uncorrelated random vectors
uniformly distributed on the unit 
circle(2D)/sphere(3D)~\footnote{Here we use this `vectorial noise' version, shown 
in~\cite{gregoire2004onset} to be less sensitive to finite size effects in the coexistence phase.
We also used the more common `angular noise' version, obtaining similar results, see~\cite{SUP}.}.
Square domains of linear size $L$ containing $N = \rho_0 L^d$ particles,
with $N$ ranging from a few million to a few billion were considered.
For numerical efficiency, small speed and weak noise were avoided. We used
$v_0=1$, $\eta=0.5$ (2D) and $0.45$ (3D) with $\rho_0=2$, parameter values in the 
homogeneous ordered phase, but not too deep inside.
Fluctuation fields $\delta\rho$ and $\delta {\bf v}_\perp$ were obtained by coarse-graining over boxes of unit linear length.
The associated correlation functions were simply obtained by computing the square norm of the fields' Fourier transform.

In finite systems with periodic boundary conditions, the direction of order diffuses slowly 
(the diffusion constant $\propto 1/N$ \cite{TT2}).
To estimate quantities scaling anisotropically like those defined by Eqs.~(\ref{eq:Space-time_CF}), 
one then needs, before averaging in time, 
to rotate a copy of the system at each measure so that global order remains along a chosen direction. 
Moreover data have then to be averaged over times longer than the timescale of this rotation.
This is possible but costly and quickly becomes prohibitive for large systems. 
Forcing global order to remain along a given direction can be achieved by either
applying an external field 
or by imposing reflecting side boundaries as in, e.g.,
\cite{Kyriakopoulos2016NJP,toner1998soundwaves,geyer2018sounds}. This perturbs slightly the global behavior of the system, but allows for 
much shorter averaging times at equivalent sizes. 
All three protocols were tested, and we found that when used cautiously
they yield identical results over the scales that can be explored by all (see \cite{SUP} for details). 
Below, we only present data obtained using a channel with reflective walls.

%%%%%%%%% Equal-time correlations %%%%%%%%%

%%%%%%%%%%%%%%%%%%%%%%%%%%%
\begin{figure}[!t]
	\includegraphics[width=0.49\textwidth]{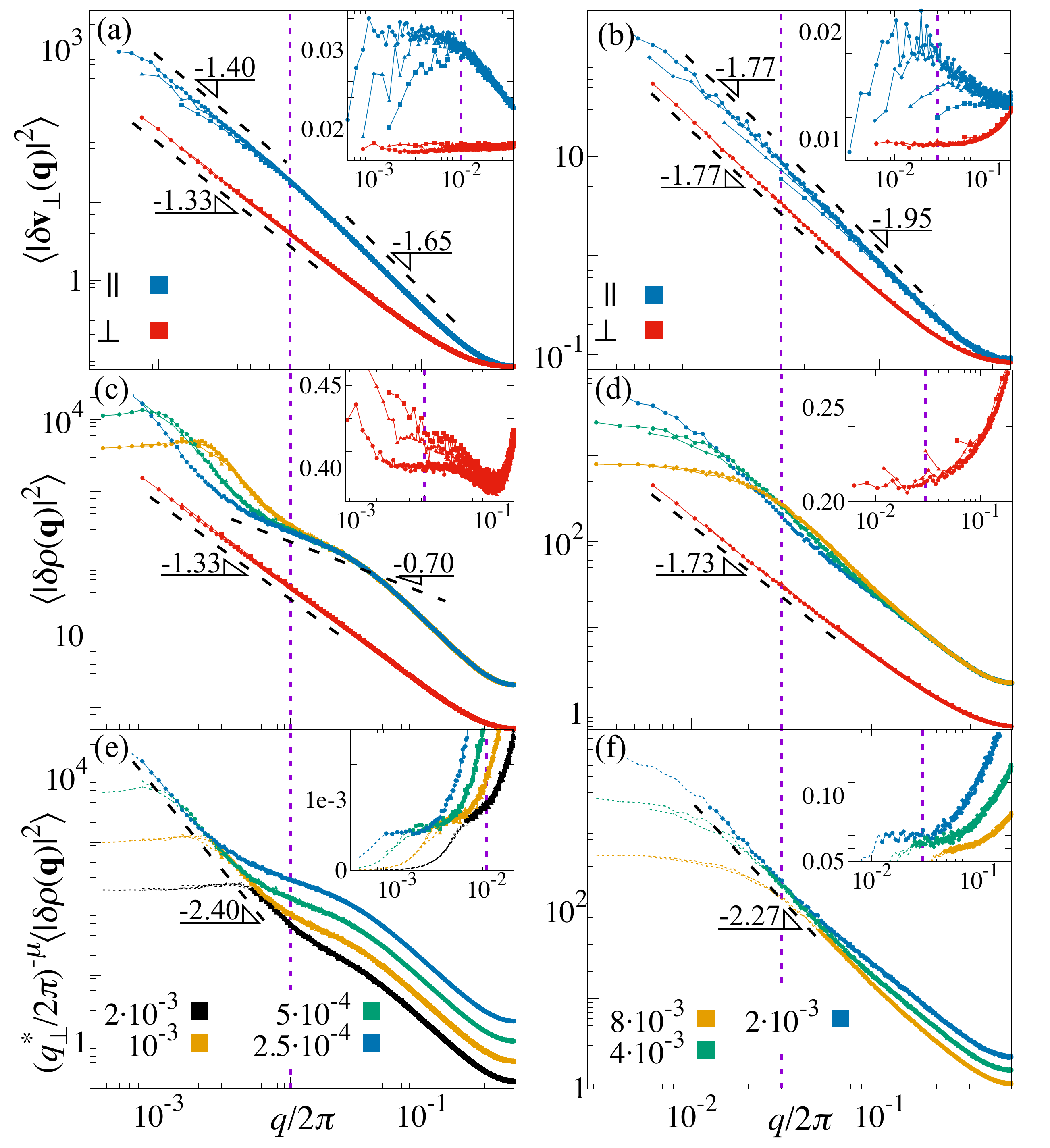}
	
	\caption{Equal time correlations in 2D (left) and 3D (right). Insets contain same data as their main panel, but rescaled
	 by $\left(\frac{q}{2\pi}\right)^{\sigma}$ where $\sigma$ is an estimated exponent. The dashed purple vertical lines mark the crossover
	 scales $q_c/(2\pi)$ (see text). Symbols code for linear system size: 
	in 2D squares, triangles and dots for $L=2000$, 4000 and 8000;
	in 3D squares, triangles, diamonds, and dots for $L=100$, 200, 500 and 960.
	(a,b): Velocity correlations in the transverse (red, lower data) and longitudinal (blue, upper data) directions.
	Insets: $\sigma = 1.33$(2D, $\perp$), $1.40$ (2D, $\|$), and $1.77$ (3D, $\perp$ and $\|$).
	(c,d): Density correlations  in the transverse (red, lower data) and longitudinal directions for different values of 
	$q^*_\perp$ (upper curves, shifted upward for clarity).
	Insets: $\sigma = 1.33$ (2D, $\perp$) and 1.73 (3D, $\perp$).
	(e,f): Longitudinal data of (c,d) rescaled by ${q_\perp^*}^{-\mu}$ with $\mu = 1$ and $\tfrac{1}{2}$ in 2D and 3D respectively.
	Points in the regime $q_{\perp} \gg q_\|$ are shown in thin dashed lines for clarity.
	Insets: $\sigma = 2.40$ (2D) and 2.27 (3D).
	}
	\label{fig:Equal-time}
\end{figure}
%%%%%%%%%%%%%%%%%%%%%%%%%%%
		
%%%%%%%%% Velocity %%%%%%%%%
Although we measured correlations in the whole ($q_\|,q_\perp$) 
plane~\footnote{In 3D we averaged over all directions of ${\bf q}_\perp$}, 
exponents $\chi$ and $\xi$ can be estimated from just the longitudinal ($q_\perp = 0$) and 
transverse ($q_\| = 0$) directions. For velocity correlations, we have: 
\begin{equation}
\label{scaling_v} 
\langle |\delta {\bf v}_\perp ({\bf q})|^2\rangle \!\underset{q\to0}{\sim}\! \left\{q_{\perp}^{-\zeta} 
\,{\rm for}\, q_{\perp}^\xi \!\gg\! q_\| {\rm ;}\;
q_{\|}^{-\zeta/\xi} \,{\rm for}\, q_\| \!\gg\! q_{\perp}^\xi  \right\}
\end{equation}
Our data in both 2D and 3D show that $\langle |\delta {\bf v}_\perp
({\bf q})|^2\rangle$ scales cleanly at small values of $q_\perp$ 
(lower sets of curves in Figs.~\ref{fig:Equal-time}(a,b)), with estimated values of $\zeta$ 
slightly but significantly
different than those conjectured by Toner and Tu (see Table~\ref{tab:Exponents}).
Behavior in the longitudinal direction is more surprising (upper set of curves in Figs.~\ref{fig:Equal-time}(a,b)).
While from~\cite{TT98} a divergence for $q_\| \to 0$ with
an exponent $-2$ is conjectured in both 2D and 3D,
we observe, in 2D, a {\it size-independent crossover} from a power law with exponent $\simeq -1.65$ 
at intermediate values of $q_\|$ to one with a {\it larger} exponent $\simeq -1.4$
at smaller $q_\|$. 
The crossover scale $\ell_c = 2\pi/q_c \simeq 100$, indicated by the purple dashed lines in our figures, 
is of the same order as typical sizes considered so far in other works~\cite{toner1998soundwaves,chate2008collective},
which may explain why it has never been reported. 
Note further that our post-crossover estimate $-1.4$ 
 is not far from the $-1.33$ value measured in the transverse direction, 
implying weak, possibly vanishing, anisotropy ($\xi\simeq 0.95$).
In 3D the two correlation functions show approximately the same exponent above a scale $\ell_c \simeq 30$: 
scaling is isotropic (Fig.~\ref{fig:Equal-time}(b)). 
Overall, our measures lead to values of $\chi$ and $\xi$ 
in clear departure from those conjectured by Toner and Tu (see Table~\ref{tab:Exponents}).

%%%%%%%%% Density %%%%%%%%%
The density correlation function is expected to show the following longitudinal and transverse scalings
\footnote{ A third intermediate scaling region with 
\unexpanded{$\langle|\delta \rho ({\bf q})|^2\rangle \underset{q\to0}{\sim}  q_{\|}^{-2} q_\perp^{2-\zeta}$} is expected for $q_\perp^\xi \gg q_{\|} \gg q_\perp$
 \cite{TT98}, but given our estimate $\xi \simeq 1$ both in 2D and 3D, we expect it to be unobservable, at odds with the earlier results of Ref.~\cite{toner1998soundwaves}.}
\begin{equation}
\label{scaling_rho} 
\!\!\!\!\langle |\delta \rho ({\bf q})|^2\rangle \!\!\underset{q\to0}{\sim}\!\! \left\{\!q_{\perp}^{-\zeta} 
\,{\rm for}\, q_{\perp} \!\gg\! q_\| {\rm ;}\; 
q_{\perp}^2 q_{\|}^{-2-\zeta/\xi} \,{\rm for}\, q_\| \!\gg\! q_{\perp}^\xi \!\right\}
\end{equation}
In the transverse direction, our data confirm that
scaling takes place with the same exponent as for velocity
correlations (Figs.~\ref{fig:Equal-time}(c,d), lower set of curves), 
albeit with more pronounced finite size effects, especially in 2D (compare insets of Fig.~\ref{fig:Equal-time}(a,b) and Fig.~\ref{fig:Equal-time}(c,d), 
see \cite{SUP} for comments). 

In 3D, the apparent exponent is slightly lower in absolute value than the one given by $\langle |\delta {\bf v}_\perp ({\bf q})|^2\rangle$ 
($-1.73$ vs.\ $-1.77$), but given the limited range of scaling available we cannot exclude that these two values are in fact the same asymptotically.

The scaling of density fluctuations in the longitudinal direction is more subtle to analyse
because it depends explicitly on $q_\perp$ (see Eq.~(\ref{scaling_rho})).
The behavior of $\langle |\delta \rho ({\bf q})|^2\rangle$ with $q_\|$ for 3 fixed values of $q^*_\perp$ 
is shown in Figs.~\ref{fig:Equal-time}(c,d) (upper sets of curves).
One can identify three regimes below the crossover scale $q_c = 2\pi/\ell_c$, which are most easily distinguished in 2D, but
probably also present in 3D.
For the smallest values of $q_\|$, the functions reach a plateau, whose range of existence and amplitude respectively increases and decreases with $q^*_\perp$.
This behavior corresponds to the ``transverse'' regime where $q_\| \ll q_\perp$.
Increasing $q_\|$ beyond this plateau, $\langle |\delta \rho ({\bf q})|^2\rangle$
shows a second scaling behavior with $q^*_\perp$-dependent amplitude, in qualitative agreement with Eq.~(\ref{scaling_rho}).
Finally, in 2D where sufficiently large systems can be studied, 
a third scaling region is observed, 
with slow (exponent $\sim-0.7$), $q^*_\perp$-independent decay whose range increases when $q^*_\perp \to 0$. 
Such a regime is absent from the framework of TT theory.
 
The second regime also departs strikingly from the Toner-Tu results.
In this region both 2D and 3D curves do not collapse when their amplitude is rescaled by 
$q^{*-\mu}_\perp$ with $\mu = 2$, as predicted exactly by TT theory,
but rather with $\mu \simeq 1$ in 2D and 0.5 in 3D (Fig.~\ref{fig:Equal-time}(e,f)).
Moreover, the collapsed curves do not decay with exponent $-2-\zeta/\xi \simeq -3.4$ (2D) and $\simeq -3.77$ (3D) 
as predicted by Eq.~(\ref{scaling_rho}) using the values of $\chi$ and $\xi$ determined
from $\langle |\delta {\bf v}_\perp ({\bf q})|^2\rangle$. Rather, we find $-2.4$ in 2D and $-2.27$ in 3D.
Our data therefore suggest that for $q_\| \gg q_{\perp}^\xi$, 
$\langle |\delta \rho ({\bf q})|^2\rangle \sim q_{\perp}^\mu q_{\|}^{-\mu-\zeta/\xi}$
with $\mu \simeq 1$ in 2D and $0.5$ in 3D. 

%%%%%%%%% Space-time correlations %%%%%%%%%

%%%%%%%%%%%%%%%%%%%%%%%%%%%
\begin{figure}[!t]
	\includegraphics[width=0.49\textwidth]{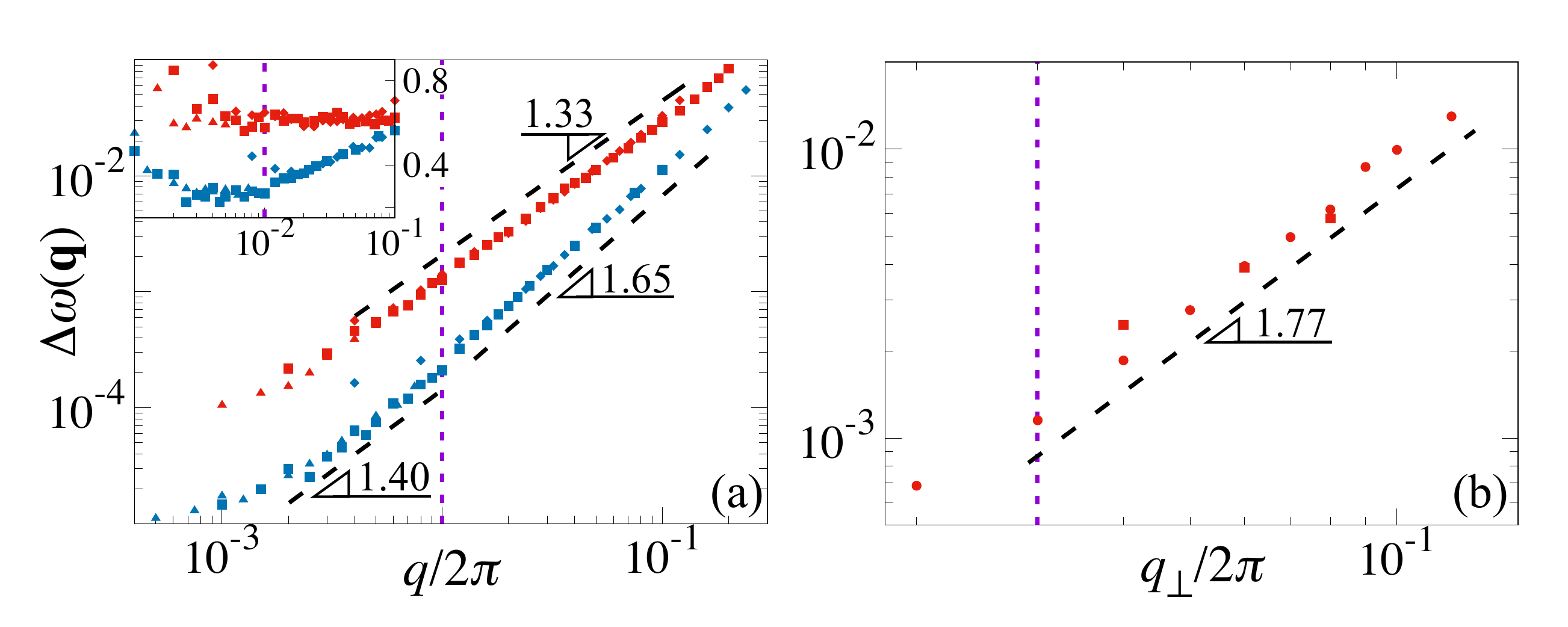}
	\caption{
	(a): 2D peak widths as functions of $q$ in the transverse (red, upper curves) and longitudinal (blue, lower curves) directions;
	insets: same data rescaled by $\left(\frac{q}{2\pi}\right)^{\sigma}$ with respectively $\sigma = 1.33$ and $1.40$.
	(Diamonds, squares and triangles respectively correspond to system sizes $L=1000$, $2000$ and $4000$.)
	(b): Same as (a) but in 3D and only in the transverse direction for the peak related to 
	${\bf v}_{\rm T}$ (see~\cite{SUP}) at sizes $L=100$ (squares) and 200 (dots).
	}
	\label{fig:Space_time}
\end{figure}
%%%%%%%%%%%%%%%%%%%%%%%%%%%

In order to assess the dynamical exponent $z$, we now turn to the study of space-time correlations.
As expected from Eqs.~(\ref{eq:Space-time_CF}) and previous work in 2D~\cite{toner1998soundwaves,geyer2018sounds}
both $\langle |\delta \rho (\omega,{\bf q})|^2\rangle$ and $\langle |\delta {\bf v}_\perp (\omega,{\bf q})|^2\rangle$,
as functions of $\omega$, show two asymmetric peaks 
that become symmetric in the transverse direction ($\theta_{\bf q} = \frac{\pi}{2}$).
In 3D, one observes the emergence of an additional third peak 
in $\langle |\delta {\bf v}_\perp (\omega,{\bf q})|^2\rangle$
coming from its component ${\bf v}_{\rm T}$.
All these peaks are well fitted close to their maximum by Cauchy distributions of the type
$H_{\pm,{\rm T}}({\bf q})/\left[1 + (\omega - \omega_{\pm,{\rm T}}^*({\bf q}))^2/\Delta\omega^2_{\pm,{\rm T}}({\bf q})\right]$, 
where $H_{\pm,{\rm T}}({\bf q})$, $\omega_{\pm,{\rm T}}^*({\bf q})$ and $\Delta\omega_{\pm,{\rm T}}({\bf q})$ respectively account for
their heights, positions and half-peak widths (see data in~\cite{SUP}).
Since we have seen that density correlations seem more sensitive to finite-size effects,
we now focus on velocity correlations for the quantitative characterization of the peaks.
As expected peak positions $\omega_{\pm,{\rm T}}^*({\bf q})$ scale linearly with $q$
in the limit $q\to 0$, and the sound speeds $c_{\pm,{\rm T}}(\theta_{\bf q})$ are given by the corresponding slopes.
Perfect agreement is found with the linear theory, both in 2D and 3D \cite{TT2}.
Peak widths, on the other hand, show non-trivial scaling:
$\Delta\omega_{\pm,{\rm T}}({\bf q})$ correspond to the dampings $\varepsilon_{\pm,{\rm T}}({\bf q})$
and thus from Eq.~(\ref{eq:scaling_damplings}) are expected to scale as $q^{z/\xi}_\|$ and $q_\perp^z$ in the longitudinal and transverse directions.
We find rather good scaling in 2D for both longitudinal and transverse
directions (Fig.~\ref{fig:Space_time}(a)), with, in this 
last case, $z\simeq 1.33$. 
In the longitudinal direction, we find weak evidence of a crossover at the same scale $\ell_c$ as for equal-time correlations.
Below $\ell_c$, the estimated value of  $z/\xi$ (1.65) is identical to that of $\zeta/\xi$ found below $\ell_c$ in
Fig.~\ref{fig:Equal-time}(a).
Beyond $\ell_c$, we unfortunately could not obtain much data, but the few points we have are compatible
with a slope 1.4, i.e. the asymptotic value of $\zeta/\xi$ found from Fig.~\ref{fig:Equal-time}(a).
In 3D, where the data is much more limited, we can nevertheless observe good scaling of the peak width
$\Delta\omega_{{\rm T}}({\bf q})$
over almost a decade in the transverse direction $\theta_{\bf q} = \frac{\pi}{2}$, yielding the estimate
$z\simeq 1.77$ (see Fig.~\ref{fig:Space_time}(b)), identical to our estimate of $\zeta$ from equal-time correlations.
Results leading to similar values of $z$ in 2D and 3D are found from the scaling of the peaks heights, see~\cite{SUP} for details. 

In summary, using the Vicsek model, 
we have measured {\it independently} the values of the three universal exponents $\chi$, $\xi$ and $z$
that characterize long-range correlations of density and velocity fluctuations in polar flocks, 
and found them incompatible with those conjectured by Toner and Tu
in~\cite{TT95} (see Table~\ref{tab:Exponents}). 
These differences indicate that at least some of the nonlinearities identified
in~\cite{toner2012reanalysis} and neglected in the original
calculation are indeed relevant asymptotically.

Our data suggest in particular the existence of a crossover scale beyond which ---{\it i.e.} at scales scarcely explored before---
there is very little or vanishing anisotropy. 
Coming back to the popular giant number fluctuations, 
we find that $\zeta/d$, which governs their scaling, varies very little across scales, 
and takes values close to those predicted by Toner and Tu, 
(see Table~\ref{tab:Exponents}, and \cite{SUP}). 
This clarifies why previous studies focusing on this quantity could not challenge the Toner and Tu conjecture~\cite{toner1998soundwaves,chate2008collective,chate2008modeling}.

We find identical estimates, within our numerical accuracy, of $\zeta$ and $z$.
In other words, the hyperscaling relation $z=d-1+2\chi+\xi$ seems satisfied.
If we take this numerical fact for granted, it implies that, somewhat counterintuitively, 
the vertices responsible for the departure from the 1995 TT results are {\it not} those coupling density and order.
Moreover, $\zeta=z$ also implies that the noise variance $\Sigma$ does not renormalize, so that the dominant effective noise
in the ${\bf v}$-equation is indeed additive, as assumed in TT theory 
(see \cite{SUP} for the simple arguments leading to these conclusions). 

We also find qualitative discrepancies with TT theory in the longitudinal behavior of density correlations \footnote{Note that our 2D 
results would be in disagreement with TT theory under the hypothesis that the three scaling regimes observed 
would correspond to those predicted (see Eq. \eqref{scaling_rho} and the intermediate one mentioned in \cite{Note3}).
In particular, the third scaling regime we identify is independent of $q_\perp$ }.
We have at present no full understanding of this, but, as explained in a forthcoming publication, 
the consideration of a (conserved) additive noise in the density equation 
---something quite natural in the context of fluctuating hydrodynamic equations--- leads to a modified form
of Eq. (\ref{scaling_rho}) in the $q_\perp\to 0$ sector. This change already occurs at the linear level 
and could account, upon renormalization, for the peculiar scaling regimes reported in Fig.~\ref{fig:Equal-time}(c-f).

All in all, our numerical results, even though they clearly rule out the Toner and Tu 1995 predictions, 
call even more than before for a complete, possibly
non-perturbative, renormalization group approach \cite{Delamotte2012NPRG}.

%\acknowledgments
We thank Cesare Nardini and Aurelio Patelli for fruitful discussions. We acknowledge a generous allocation of cpu time on Beijing CSRC's Tianhe supercomputer.

\bibliography{Toner_Tu_v7.bib}

\end{document}